\documentclass[aps,prl,twocolumn,showpacs,groupaddresses,superscriptaddress]{revtex4}
\usepackage{bbm,amssymb,amsmath,amsbsy,times,pifont,pxfonts}
\usepackage[dvips]{graphicx}
\usepackage{hyperref}
\usepackage[latin1]{inputenc}
\usepackage{subfigure}
\usepackage{fancyhdr,makeidx}
\usepackage{color}

\renewcommand{\prl}{{\it Phys. Rev. Lett.} }
\renewcommand{\pra}{{\it Phys. Rev. A} }

\newcommand{\N}{\cal N}

\newcommand{\id}{\mathbbm{1}}

\newcommand{\gr}[1]{\boldsymbol{#1}}
\newcommand{\be}{\begin{equation}}
\newcommand{\ee}{\end{equation}}
\newcommand{\bea}{\begin{eqnarray}}
\newcommand{\eea}{\end{eqnarray}}
\newcommand{\ket}[1]{|#1\rangle}
\newcommand{\bra}[1]{\langle#1|}

\newcommand{\sig}{\gr{\sigma}}

\newcommand{\eq}[1]{Eq.~(\ref{#1})}

\newcommand{\lup}[1]{\lambda^{\uparrow}_{#1}}
\newcommand{\ldn}[1]{\lambda^{\downarrow}_{#1}}

\newcommand{\an}[1]{\left\langle{#1}\right\rangle}

\newtheorem{proposition}{Proposition}
\newcommand{\proofend}{\hfill\fbox\\\medskip }

\begin{document}

\title{{Maximal Gaussian entanglement achievable by feedback controlled dynamics}}

\author{Alessio Serafini}
\affiliation{Department of Physics \& Astronomy, University College London, 
Gower Street, London WC1E 6BT, United Kingdom}

\author{Stefano Mancini}
\affiliation{Dipartimento di Fisica, Universit\`a di Camerino, I-62032 Camerino, Italy}

\date{\today}

\begin{abstract}



We determine a general upper bound for the steady-state entanglement achievable 
by continuous feedback for a system of any number of bosonic degrees of freedom.
We apply such a bound to the specific case of parametric interactions -- the most common practical 
way to generate entanglement in quantum optics -- and point out optimal feedback strategies that achieve the 
maximal entanglement. We also consider the case of feedback schemes entirely restricted to local  
operations and compare their performance to the optimal, generally non-local, schemes.

\end{abstract}


\maketitle

The field of quantum control is central 
in the current rise of quantum technologies \cite{wisebook,expi}. 
In particular, the control of the coherent resources of quantum states  
is an issue of major interest. 
Most valuable, and delicate, among such resources is certainly quantum entanglement,
whose control is a primary requisite for quantum information and 
communication \cite{theori,manciowise,passive,hill}.
This paper addresses the question of how much entanglement can be 
generated by controlling the dynamics of a bosonic quantum system,
and leads to the determination of optimal control schemes 
-- achieving maximal 
entanglement -- in relevant practical cases.
In particular, we will consider systems subject to generic quadratic Hamiltonians and losses, 
and derive a bound on the maximal entanglement achievable,
between specific bipartitions, 
by feedback schemes
based on general continuous measurements and linear driving \cite{WisMilFeedback}.
The class of dynamics and feedback strategies covered in our study 
is very important in quantum optics, and is applicable to more general 
continuous variable systems (ranging from atoms to nano-mechanical 
resonators). 
Being crucial for the implementation of a number of quantum information protocols
\cite{braunvl}, the optimisation of the generation of continuous variable entanglement
has been drawing considerable attention in recent years \cite{theori,manciowise,passive}.  
Since entanglement is {\em not} a
linear figure of merit in the quantum state's parameters, one cannot 
tackle this optimisation with standard tools, 
like semi-definite programming \cite{wido05}, but rather requires the more 
detailed, specific analysis we shall present.

\noindent {\em Notation --} We consider systems of $N$ degrees of freedoms
described by pairs of canonical operators: defining a
 vector of operators 
$
 \hat{\mathbf{x}}=\left( \hat q_{1},\hat
   p_{1},...,\hat q_{N},\hat p_{N}\right) ^\top 
$,  
 one has $\left[\hat x_{j},\hat x_{k}\right] =i  \Omega _{jk}$,
 where $\Omega$ is the $(2N)\times(2N)$  symplectic form:
$\Omega_{jk} = \delta_{j+1,k}[1-(-1)^j]/2 -
 \delta_{j,k+1}[1+(-1)^j]/2$, in terms of Kronecker deltas $\delta_{j,k}$.
Also, $\hat a_j=(\hat{q}_j+i\hat{p}_j)/\sqrt{2}$.

For a system with such a phase-space structure we can define 
``Gaussian states'' as the states with Gaussian Wigner functions. 
These states are completely determined by the vector of means 
$\an{\hat{\bf x}}$, and by the covariance matrix (CM) $\sig$, 
with entries
 $\sigma_{jk}=( \langle \Delta \hat x_{j}\Delta \hat
    x_{k}\rangle +\langle \Delta \hat x_{j}\Delta \hat x_{k}\rangle
  )$, where $\Delta \hat{o}=(\hat{o}-\an{\hat{o}})$ 
  for operator $\hat{o}$.
  The -- always necessary -- Robertson-Schr\"odinger uncertainty relation 
  is also sufficient for Gaussian states to be physical \cite{Hol75}: 
\begin{equation}
\sig + i  \Omega \geq 0 \; .
\label{heis} 
\end{equation}

We will consider Hamiltonians $\hat H$ that are at most of the second-order in $\hat{\bf x}$, 
so that their resulting free evolutions are affine in phase-space:
$
\hat{H}=(1/2)\hat{\mathbf{x}}^\top H\hat{\mathbf{x}} -\hat{\mathbf{x}}^\top \Omega B\mathbf{u}(t)
$,
where the ``Hamiltonian matrix'' $H$ is real and symmetric and $B$ is real.
The second term of 
$\hat H$ is a `linear driving' proportional to
a time-dependent input ${\bf u}(t)$: 
this term will describe the control exerted over the system.  

The system is considered to be open and such that each degree of freedom has its own channel to interact with the environment.
Though thermal noise can also be treated along the lines we will present here, 
in this study we specialise for simplicity to pure losses,
which are the main source of decoherence in quantum optical settings.
We will thus assume a beam splitter-like (``rotating wave'') 
interaction 
between each mode and the associated modes of the bath.
Under the conditions set out above,
the first moments of the canonical operators evolve according to  
${{\rm d}\langle \mathbf{\hat{x}}\rangle}/{\rm d}t  =  A\langle \hat{\mathbf{x}}\rangle +B\mathbf{u}(t)$,
while the second moments obey
\be
{\rm d}{\sig}/{\rm d}t  = A\sig+\sig A^\top + \id. \label{dsigmadt}
\ee
Here,  $A = (\Omega H - \id)/2$ is the ``drift matrix'', and $\id$ stands for the identity matrix 
with dimension clear from the context. 
We will only address stable systems, for which $(A+A^{\sf T}) < 0$.  
Note that, for Gaussian states, these equations describe the complete dynamics of the system.

As customary in the context of feedback control, we will now assume 
that the degrees of freedom of the environment can be continuously monitored on time-scales which 
are short with respect to the system's response time \cite{diosiwis}. 
{The most general (efficient) measurement on the environment}
with outcomes continuous in time 
corresponds to monitoring the operators
$(\hat{\bf a}^\top \id + \hat{\bf a}^\dag \Upsilon)$, 
where the vector $\hat{\bf a} = (a_1,\dots,a_N)^\top$ 
contains all the annihilation operators of the system, 
and the complex matrix $\Upsilon$ parametrises the measurement.
These measurements (also known as `general dyne' detections, 
see \cite{wisebook}) 
are very general, including heterodyne and 
homodyne detections as special cases, and define 
the broad setting of ``continuous feedback'' \cite{wisebook,WisMilFeedback}.
See \cite{wisebook} for a description of the POVM giving rise to such measurements.
In turn, $\Upsilon$ defines the so called ``unravelling matrix'' $U$,
given by
\begin{equation}
U \coloneqq  \frac{1}{2}\left( 
\begin{array}{cc}
\id+\text{Re}\left[ \Upsilon \right]  & \text{Im}\left[ \Upsilon\right]  \\ 
\text{Im}\left[ \Upsilon\right]  & \id-\text{Re}\left[ \Upsilon\right]
\end{array}
\right).
\end{equation}
The only conditions on $\Upsilon$ are that $U$ be symmetric and 
positive semi-definite.
The outcome of the measurements on the environment is recorded as a ``current'' 
${\bf y} = C\langle \mathbf{\hat x}\rangle +\frac{d\mathbf{w}}{dt}$,
where $C = 2 U^{1/2} \bar{C}$ and $\bar{C}_{jk}= (\delta_{2j-1,k}+\delta_{2(j-N),k})/\sqrt{2}$
for $j,k\in[1,\ldots,2N]$. Finally, ${\rm d}\mathbf{w}$ is a vector of real Wiener increments satisfying 
${\rm d}\mathbf{w}{\rm d}\mathbf{w}^\top =\id {\rm d}t$ \cite{wisebook}. 
Clearly this treatment, like any feedback model, applies to systems where the 
output channels are open to experimental scrutiny like, {\em e.g.}, 
light modes resonating in a cavity (where leaking light can be detected). 
The {\em conditional} evolution 
of the moments under such continuous measurements can be derived 
by standard techniques (It\^o calculus).
It amounts to a diffusive equation with a stochastic component for the first moments $\an{\hat{\bf x}}$, 
and to a {\em deterministic} Riccati equation for the second moments \cite{wido05}. 
In our reasonings to follow, we will not make use of the details of such equations directly.
We will be interested in stable systems, and 
will determine the maximal entanglement achievable at steady state. 
Hence, all we need to remark is that a CM $\sig$  
is a {\em stabilising solution} \cite{Zhou96} of the Riccati equation for the second moments 
if and only if \cite{wido05}: 
\be
A\sig + \sig A^{\sf T} +\id \ge 0 \; . \label{ss}
\ee
Together with Inequality (\ref{heis}), this relationship completely determines the set of 
stabilising solutions of our conditional dynamics.

The final ingredient of the dynamics 
is the dependence of the linear driving
${\bf u}(t)$ on the history of the measurement record ${\bf y}(s)$ for $s<t$,
which affects both first and second moments of the {\em unconditional},
`average', evolution 
(whereas the second moments of the {\em conditional} states 
are unaffected by the linear driving), 
and closes the control loop.
We will denote the unconditional state by $\varrho$. 
Note that, for our class of dynamics, 
$\varrho$ is a statistical mixture of states with the same 
conditional CM $\sig$, obeying Inequality (\ref{ss}), and varying first moments. 
For Gaussian states, this implies that $\varrho$ can be obtained from a Gaussian state $\varrho_{0}$ with CM 
$\sig$ and vanishing first moments by local operations and classical communication alone: 
$\varrho = L(\varrho_{0})$, where $L$ is some LOCC map.

The typical aim of control over some time interval is to optimise the expected value of a {\em cost function}
\cite{wisebook, Zhou96}. 
Our cost function will be the entanglement of Gaussian multi-mode steady states for 
bipartitions of $1$ versus $(N-1)$ modes and `bi-symmetric' bipartitions ({\em i.e.,} invariant under the permutation of 
local modes). Such an entanglement can be quantified by the logarithmic negativity 
$E_{\N}=-\log_{2} \tilde{\nu}_{-}$, where $\tilde{\nu}^2_{-}$ 
is the smallest eigenvalue of $(-\sig \tilde{\Omega}\sig 
\tilde{\Omega}^{\sf T})$, $\tilde{\Omega}$ being the 
partial transposition of $\Omega$ \cite{PT,serafozzi06}.
Clearly, $\tilde{\nu}_{-}$ is {\em not} a quadratic cost function ({\em i.e.}, it is not linear in $\sig$).
This is why, albeit dealing with linear systems with Gaussian noise, 
we cannot resort to optimisation methods borrowed from classical LQG control problems \cite{wido05}. 



\noindent {\em General results --}
The main analytical result of this paper is presented 
here. 
{Its proof may be found in appendix.}

\begin{proposition}[Maximal entanglement]\label{entb}
Let $\varrho$ be a steady state achievable  
by continuous Gaussian measurements 
and linear driving for a system of any number 
of bosonic modes subject to losses and to a Hamiltonian matrix $H$.
The logarithmic negativity $E_{\N}(\varrho)$ of any $1$ versus $(N-1)$ modes 
or bi-symmetric bipartition of $\varrho$ is bounded by: 
\be
E_{\N} (\varrho) \le \max\left[0,-\frac12\log_2{(\alpha^{\uparrow}_{1}\alpha^{\uparrow}_{2})}\right] \; , \label{entbb}
\ee
where $\{\alpha^{\uparrow}_{j}\}$ are the (strictly positive) eigenvalues of $(-A-A^{\top})$ 
in increasing order, and $A=\frac12(\Omega H-\id)$. 
\end{proposition}

Inequality (\ref{entbb}) corresponds to 
\be
\tilde{\nu}_{-}^{2} \ge \alpha^{\uparrow}_{1}\alpha^{\uparrow}_{2} \; , \label{feedbound}
\ee
in terms of the smallest partially transposed symplectic eigenvalue of the 
Gaussian state $\varrho$.


The bound above applies to both conditional and unconditional 
states. In practice, only unconditional states are of interest since, 
although the first moments of the conditional states are in principle known, 
they fluctuate so fast (on the time-scale of the environment's dynamics) that 
the actual experimental state is the unconditional, average one. 
This is where the linear driving plays its crucial role in preserving the entanglement. 
Since the entanglement (for us, the logarithmic negativity) only depends
on the second moments and decreases under LOCC, 
and since the second moments of the conditional states do not depend on the linear drive,
the optimal choice for the 
linear driving is the one, always existing, that keeps the first moments fixed (say, at zero). 
In this way, the linear drive's action guarantees that the unconditional state is at all times 
a conditional state -- satisfying Inequality (\ref{ss}) -- with vanishing first moments.
Hence, the optimal entangling strategy only depends on the optimal
unravelling matrix $U$. 
\noindent {\em Applications --} 
Our theoretical result applies in general to all bosonic systems subject to losses and 
quadratic Hamiltonians.
Here, we focus on optical modes oscillating in a damped cavity 
and interacting through a parametric $\chi^{(2)}$ crystal or more general nonlinear media 
(a ``non-degenerate, multi-frequency optical parametric oscillator'' \cite{comb}). 
Parametric interactions are the state of the art technology to generate 
continuous variable entanglement. 
Also, optical bosonic systems
can be interfaced with atomic systems \cite{julsgaard},
so that the feedback 
scheme
could be used to control atomic entanglement as well.


The parametric interaction between modes $j$ and $k$ is described
by the Hamiltonian $\chi(\hat{q}_j\hat{p}_k+\hat{p}_j\hat{q}_k)$ \cite{reid}. 
We will assume equal interaction strengths $\chi\ge0$ 
between each pair of modes, consider a $(n+n)$-mode bipartition, 
and describe analytically the scaling of the control of the entanglement 
with the number of modes $n$ (we also define $N=2n$). 
Our bound in this case is tight, and yields the actual 
optimal entanglement achievable by continuous filtering.
Due to the symmetry of the system
under the exchange of any two modes, the entanglement between the 
$n$-modes subsystems can be reduced to two-mode entanglement 
\cite{serafozzi04}: a local 
symplectic transformation exists that turns the matrix $A$ into an equivalent two-mode drift matrix $\bar{A}$, plus 
a direct sum of irrelevant decoupled single-mode matrices 
The matrix
$\bar{A}$ reads:
\be
\bar{A} = \left( \begin{array}{cccc}
(n-1)\chi & 0 & n\chi & 0 \\
0 &  -(n-1)\chi & 0 & -n\chi \\
n\chi & 0 & (n-1)\chi & 0 \\
0 & -n\chi & 0 & -(n-1)\chi \\
\end{array}
\right) -\frac{\id}{2} . \label{paradrift}
\ee
For the system to be stable one must require: $\chi < \frac{1}{2(N-1)}$
(unstable systems, although in principle capable of generating substantial entanglement, 
are in practice not controllable and certainly undesirable).
As $\bar{A}$ is symmetric and invertible, the `free' steady state 
CM ${\sig_{f}}$ can be promptly determined from \eq{dsigmadt}:
$\sig_{f} = -\bar{A}^{-1}/2$. 
Its logarithmic negativity is given by $\frac12\log_2[(1+2\chi)(1+2(N-1)\chi)]$.
Instead, the bound of Inequality (\ref{entbb}) for any steady state CM $\sig$ 
with continuous feedback control reads 
\be
E_{\N} \le -\frac12 \left(\log_2(1-2\chi)+\log_2\left[1-2(N-1)\chi\right]\right) \; . \label{parabound}
\ee
This upper bound is attained by the CM 
$\sig_{opt}=R^{\sf T}{\rm diag}({\alpha}_{2},1/{\alpha}_{2},1/\alpha_{1},{\alpha}_{1})R$, 
where $R$ is the orthogonal transformation that diagonalises $\bar{A}$ 
and $\{{\alpha}_{j}\}$ are the eigenvalues of $-2\bar{A}$ in 
increasing order. This solution also saturates the 
Inequalities (\ref{ss}) and (\ref{heis}).
Both the free asymptotic entanglement and the optimal one under 
continuous filtering have thus been obtained analytically. 
Once the optimal achievable state is known as is the case here,
the ``optimal unravelling'' $U_{opt}$, and hence the optimal feedback 
scheme, can be straightforwardly derived since
$U_{opt}=E(\bar{A}\sig_{opt}+\sig_{opt}\bar{A}^{\sf T} + \id)E^{\sf T}$,
where $E=(2\bar{C}\sig_{opt}-\bar{C})$ \cite{wido05}.
For two modes, this rigorously proves that the schemes considered in Ref.~\cite{manciowise} 
are indeed optimal.

%
%

\begin{figure}
\begin{center}
\includegraphics[width=0.4\textwidth]{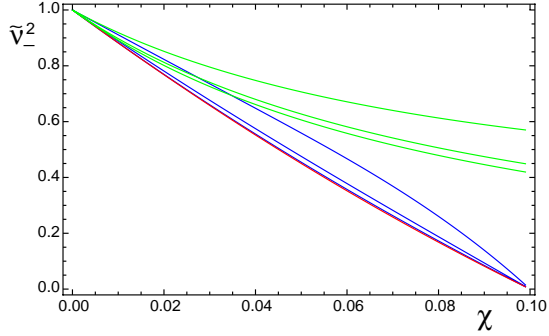}
\end{center}
\vspace{-0.5cm}
\caption{\label{fig} 
Squared symplectic eigenvalue $\tilde{\nu}_{-}^2$ at steady state for a system of 6 modes 
($\tilde{\nu}_{-}\rightarrow0$ implies infinite entanglement).
Green (lighter) curves depict $\tilde{nu}_{-}^{2}$ in 
the absence of control (from top to bottom: 1:5, 2:4, and 3:3 modes bipartition); bleu (darker) curves refer to 
numerically optimised local feedback (from top to bottom: 1:5, 2:4, and 3:3 modes bipartition);
the red curve is the analytical lower bound (\ref{feedbound}).}
\end{figure}

\noindent {\em Local control -- } Such an optimal entanglement 
is in general achieved by filtering the system through {\em global} measurements 
on the environment, as no restrictions were assumed for the unravelling matrix $U$.
This applies to situations where 
the output channels of the two local subsystems 
can be combined before being measured (like, {\em e.g.}, for a parametric crystal in a cavity).
We intend now to provide a lower bound on the entanglement achievable under {\em local} control,
where the environmental degrees of freedom pertaining to the separate subsystems 
cannot be combined, and compare it to the upper bound we obtained above. 
To this end, we will adopt direct (Markovian) feedback \cite{WisMilFeedback} and set 
$
\mathbf{u}(t)  = F \mathbf{y}(t)
$.
The unconditional 
evolution of the system is then described by
\bea
{\rm d}{\sig}/{\rm d} t =A^{\prime} {\sig}+{\sig} A^{\prime \sf T} +D^{\prime} \, ,
\label{coveq}
\eea
with drift and diffusion matrices modified as 
$A^{\prime}=\bar{A}+BFC$ and 
$D^{\prime}=\id-C^{\sf T}F^{\sf T}B^{\sf T}-BFC+2BFF^{\sf T}B^{\sf T}$.
We also choose a specific form of $U$ and $BF$. 
Since in the free dynamics,  
governed by the drift matrix of \eq{paradrift}, the quadratures 
$\hat p_1$ and $\hat p_2$ are less noisy than $\hat q_1$ and $\hat q_2$, 
it is advantageous to monitor locally $\hat p_1$ and  $\hat p_2$ and drive with the respective currents the quadratures $\hat q_2$ and $\hat q_1$. 
However, due to the possible asymmetry of the two subsystems for $m \neq n$, 
we have to consider different driving amplitudes $\mu_1$ and $\mu_2$ for their quadratures.
All this corresponds to setting $U_{33}=U_{44}=1$, $\sqrt{2}(BF)_{24}=\mu_2$,
$\sqrt{2}(BF)_{43}=\mu_1$,
and all other entries of $U$ and $BF$ vanishing.
We can then find the steady state solution of \eq{coveq}
as a function of the two feedback amplitudes $\mu_1$ and $\mu_2$, 
and evaluate its logarithmic negativity.
It turns out that the maximum logarithmic negativity at steady state is attained for $\mu_2=\mu_1 n/m$. 
Hence, we are left with the entanglement depending on one parameter, 
over which we minimise numerically in
the stable region, determined by $(A^{\prime}+A^{\prime\sf T})<0$. 
As a case of study, we have considered a system of 6 modes and summarised the results in 
Fig.~\ref{fig}.
Because of the symmetry of the Hamiltonian, 
local control is very close to optimal global control in the case of a balanced bipartition.
However, the more unbalanced the bipartition, the more degraded the control, although numerics indicate that 
arbitrarily large entanglement can always be retrieved
approaching the instability.

Before concluding, let us further emphasise the usefulness of feedback control 
by describing the practical case of two modes with interaction strength to loss factor ratio 
$\chi=0.45$. 
Without control, this system would generate $0.93$ ebits of logarithmic negativity at steady state. 
The optimal feedback control would rise this value to $3.32$ ebits.  
The Markovian local control discussed here, instead, allows one to reach $2.12$ ebits: 
a remarkable improvement over 
the case with no control showing that, in this instance, 
about half of the entanglement retrievable by global measurements 
can be recovered from the environment by local measurements. 


\noindent {\em Conclusion -- } We derived a bound on the entanglement 
achievable, at steady state and for various bipartitions, 
in multimode linear bosonic systems under continuous feedback control.
When applied to the practical case of symmetric parametric interactions,  
our bound also allows one to determine the 
measurement strategy maximising the steady-state entanglement, 
which is relevant to optimise the experimental generation of 
continuous variable entanglement, and hence useful 
for countless quantum information protocols \cite{braunvl}.
More generally, 
our investigation yields a technique for the optimisation of nonlinear figure of merits in bosonic quantum systems, with a broad range of applications in
quantum information processing and state engineering.


\noindent {\em Appendix -- {Proof of Proposition 1.}} 
{Henceforth, $\ket{v}$ will stand for a unit vector in the phase space $\Gamma$ 
and
$\{\ \lup{j} \}$ ($\{\ \ldn{j} \}$) will be 
the $2N$ increasingly-ordered (decreasingly-ordered)
eigenvalues of an $N$-mode CM $\sig$.
For each $\ket{v}$, one can define the unit vector $\ket{w} = \tilde{\Omega}\sig^{1/2}\ket{v}/
\sqrt{\bra{v}\sig\ket{v}}$, such that $\bra{v}\sig^{1/2}\ket{w}=0$ (
since $\tilde{\Omega}=-\tilde{\Omega}^{\sf T}$) and 
{\be
\tilde{\nu}^{2}_{-} 
\ge\min\;
\bra{v}\sig\ket{v}\bra{w}\sig\ket{w} = \lup{1}\lup{2} \, . \label{sympb}
\ee
with the $\min$ taken over $\ket{v}, \ket{w}$ satisfying 
$\bra{v}\sig^{1/2}\ket{w}=0$.}


{We will further denote by $\ket{v_j}$ the eigenvectors corresponding to the 
increasingly ordered eigenvalues of $\sig$: $\sig\ket{v_j} = \lup{j}\ket{v_j}$.
Then, by using the Robertson Schr\"odinger Inequality and the `Poincar\'e Inequality' \cite{bathia},
one can show that 
 a vector $\ket{w}$ must exist in $\Omega\Gamma_k$ (defined as the subspace spanned by the $k$ 
 orthogonal vectors $\Omega\ket{v_k}$)
for which $\bra{w}\sig\ket{w} \le \ldn{k} $, and such that
\be
\lup{k}\ldn{k}\ge 1. \label{eigunc}
\ee
Now, let $\sig_{\infty}$ be a conditional CM at steady state obtained under 
continuous measurements, pure losses and a 
Hamiltonian matrix $H$. 
Applying Inequality (\ref{ss})
to the eigenvectors corresponding to $\ldn{1}$ and $\ldn{2}$, one has
for the two largest eigenvalues $\ldn{1}$ and $\ldn{2}$ of $\sig$:
\be
\ldn{1}\ldn{2} \le \frac{1}{\alpha^{\uparrow}_{1}\alpha^{\uparrow}_{2}} \; ,
\label{eigb}
\ee
where $\{\alpha^{\uparrow}_{j}\}$ are the (strictly positive) eigenvalues of $(-A-A^{\top})$ 
in increasing order. 
The chain of Inequalities (\ref{sympb}), (\ref{eigunc}) and (\ref{eigb}) leads to 
(\ref{feedbound}) for the partially transposed 
symplectic eigenvalue of the conditional state.}

{Finally, as we have seen previously, 
$\varrho=L(\varrho_0)$, where $L$ is a LOCC operation and 
$\varrho_{0}$ a Gaussian state with a CM which is a stabilising solution of 
(\ref{dsigmadt}). Hence 
$E_{\N}(\varrho)=E_{\N}(L(\varrho_0)) \le E_{\N}(\varrho_0)\le \max\left[0,-\log_{2}(\alpha_{1}^{\uparrow}\alpha_{2}^{\uparrow})/2\right]$,
where (\ref{feedbound}), the formula $E_{\N}=-\log_2(\tilde{\nu}_{-})$,
and the monotonicity of $E_{\N}$ 
under LOCC \cite{plenioln} have been invoked.
\proofend}


\noindent{\em We acknowledge financial support from the EU through the 
FET-Open Project HIP (FP7-ICT-221899).}



\begin{thebibliography}{99}
\bibitem{PT}Partial transposition on a $(m+n)$-mode bipartition is described in phase space by 
$T=\id \oplus \sigma_z^{\oplus n}$, where the identity act on the first $m$ modes while the direct sum of reflections $\sigma_{z}$ acts on the other $n$ modes. 
One then has $\tilde{\Omega}=T\Omega T$.
\end{thebibliography}
\end{document}